\documentclass[twocolumn, aps, pra, 10pt, superscriptaddress, floatfix, nobibnotes]{revtex4-1}
\usepackage{amssymb}
\usepackage{amsmath}
\usepackage{epsfig}
\usepackage{graphics, graphicx}
\usepackage{bbold}
\usepackage{psfrag}
\usepackage{mathcomp}
\usepackage{subfigure}
\usepackage{verbatim}
\usepackage{bm}
 
\usepackage{color}

\def\cp#1{\mathbf{#1}}
\usepackage[hidelinks,hypertexnames=false]{hyperref}
\makeatletter
\def\frontmatter@RRAP@format{\small\centering}
\makeatother

\begin{document}

\title{Interaction-induced dimension reduction for bound states in microwave-shielded ultracold molecules}
\author{Haitian Wang}\thanks{{These authors contributed equally.}}
\affiliation{Beijing National Laboratory for Condensed Matter Physics, Institute of Physics, Chinese Academy of Sciences, Beijing, China}
\affiliation{School of Physical Sciences, University of Chinese Academy of Sciences, Beijing, China}
\author{Tingting Shi}\thanks{{These authors contributed equally.}}
\affiliation{Beijing National Laboratory for Condensed Matter Physics, Institute of Physics, Chinese Academy of Sciences, Beijing, China}
\author{Xiaoling Cui}
\email{xlcui@iphy.ac.cn}
\affiliation{School of Physical Sciences, University of Chinese Academy of Sciences, Beijing, China}
\date[]{(Dated: September 24, 2026)\\ \mbox{}}

\begin{abstract}
{Microwave-shielded ultracold molecules provide a powerful platform for exploring quantum physics driven by long-range interactions. However, few-molecule bound states in fully three-dimensional (3D) environments remains largely unexplored.} Here we show that the tetratomic and hexatomic bound states of 3D ultracold molecules dressed by a single elliptic microwave field can be accurately described by effective one-dimensional (1D) models incorporating high-order angular fluctuations. We identify the validity region of such 1D description in the parameter plane of microwave field ellipticity and coupling strength. The hard-core character of 1D models enables a duality between bosonic and fermionic molecules in real and spectral space, while their momentum distributions remain distinct. Our results demonstrate an effective dimension reduction purely due to the intrinsic interaction anisotropy rather than any external confinement.  Extending to large systems, our results suggest a self-bound single-molecule array as the ground state of both bosonic and fermionic molecular gases.
\end{abstract}
\maketitle

{
\section*{Introduction} }

Few-body physics plays an indispensable role in ultracold gases, where the formation of few-body bound states governs both the practical stability of the system and the emergence of many-body phases. For instance, the existence of Efimov trimers\cite{Efimov1,Efimov2}, which exhibit discrete scaling invariance and binding energies sensitive to a short-range three-body parameter, typically induces significant three-body loss in atomic gases\cite{review_PR,review_RMP,review_RPP}.
In contrast, a distinct class of universal bound states exists, whose properties are independent of short-range details and are determined by only a few physical parameters\cite{KM, Blume, Petrov,  Pricoupenko, Parish,Cui, KM_1D,Mehta_1D,Petrov_1D}. These clusters are expected to be stable against inelastic collisions and are responsible for universal many-body phases with intriguing higher-order correlations\cite{Parish3, Parish4, Naidon, mass_polaron, QSF}.
While all these bound states have been extensively investigated in atomic systems with short-range interactions, their counterparts in long-range interacting systems remain largely unexplored.

As an ideal platform for simulating long-range interactions, ultracold dipolar molecules have recently gained rapid developments thanks to the technique of microwave shielding\cite{Hutson, Quemener, Doyle, Luo1, Luo2, Luo3, ZhangJ, Wang, Will1, Will2, Will_new,Wang2}. This method creates a large repulsive shielding core (on the order of $100$-$1000\ a_B$, where $a_B$ is the Bohr radius), which prevents molecules from approaching closely and thus strongly suppresses inelastic two-body losses. These developments have enabled milestone achievements such as Fermi degeneracy\cite{Luo1} and Bose-Einstein condensation\cite{Will2, Will_new, Wang2} of molecular gases.
Furthermore, by tuning the ellipticity and coupling strength of the microwave field, experiments have successfully observed scattering resonances\cite{Luo2}, tetratomic bound states\cite{Luo3, ZhangJ} in fermionic molecules, and self-bound droplets in bosonic systems\cite{Will_new,Wang2}. Theoretically, research has predominantly focused on two-body scattering properties\cite{Shi3, Hutson2, Zhang, Hutson4, Xu} and  bound states\cite{Hutson4, Bohn, Bohn2, Croft}, as well as many-body fermionic superfluids\cite{Shi3} and bosonic droplets\cite{Shi1,Shi2,Langen1,Langen2}. However, the exploration of fundamental few-body physics in dipolar systems has only been at its infancy\cite{Greene1, Greene2, DWWang, Endo, Greene, SWC}.
Drawing on the established importance of few-body physics in atomic systems, a comprehensive understanding of its role in dipolar molecules is highly demanded, for both precisely controlling collisional stability and guiding the search for novel many-body phases therein.

A number of previous studies have explored few-body physics in long-range interacting dipolar systems. For example, the Efimov effect and universal bound states have been studied in  bosonic\cite{Greene1,Endo} and fermionic\cite{Greene2, Endo} systems, taking an interaction combining  isotropic short-range and anisotropic long-range potentials. However, such interaction potential differs significantly, in both its long- and short-range behavior, from the microwave-shielded potential relevant to ultracold molecules.
Adopting a similar shielding potential, Ref.\cite{DWWang} has studied the bound state formation in { two-dimensional (2D)} molecules.  Recently, as the first attempt of few-body problem in { three-dimensional (3D)} microwave-shielded molecules, we considered the case of extreme interaction anisotropy under a highly elliptic microwave field\cite{SWC}. There, effective one-dimensional (1D) models incorporating the lowest-order angular fluctuations were established to describe universal bound states, revealing a strong suppression of Efimov physics and the emergence of a Bose-Fermi duality.
Nevertheless, in realistic experiments of ultracold molecules, microwave fields are generally with small ellipticity, as this is essential for a strong shielding effect in suppressing two-body losses\cite{Huston5}.  For this practical regime, it remains an open question whether the effective models and bound state properties identified previously\cite{SWC} are still applicable.

In this work, we investigate the bound states of ultracold molecules under realistic microwave shielding conditions, specifically focusing on small ellipticities. For this regime, we develop an improved effective 1D model that incorporates high-order angular fluctuations. This model yields accurate predictions for both the tetratomic binding energy and the Born-Oppenheimer potential in  three-molecule sector.
By comparing our model with exact solutions of tetratomic bound states, we identify its validity regime in the parameter plane of ellipticity and coupling strength of microwave field. Encouragingly, we find 1D description remains valid even at small ellipticities for typical microwave coupling strengths on the order of tens of MHz.
Within this valid regime, the Bose-Fermi duality in real and spectral space is preserved. The quantum statistics of the molecules, however, remain distinguishable through their momentum distributions. Furthermore, we find that hexatomic molecules are generally more deeply bound than tetratomic ones across a wide range of ellipticities and microwave couplings. For the ground state, the hexatomic bound state is well-approximated by two linked tetratomic states. This structure, when extended to larger systems, suggests the formation of a self-bound ``single-molecule array" for both bosonic and fermionic molecules.

Our work demonstrates an improved effective 1D model to accurately describe the bound state formation in 3D ultracold molecules. Crucially, this dimensional reduction stems purely from the intrinsic anisotropy of interaction potential, a mechanism qualitatively distinct from conventional method of external confinement. The resulting universal few-body physics is fundamentally different from those in short-range interacting atomic systems and should be directly observable in current experiments with ultracold molecules.

{
\section*{Results and Discussion}
\subsection*{Interaction model} }

We consider ultracold molecules dressed by an elliptic microwave field ${\cp E}= Ee^{i(kz-\omega t)}({\cp e}_+\cos\xi  +{\cp e}_-\sin\xi)+c.c.$, where ${\cp e}_{\pm}=\mp ({\cp e}_x\pm i{\cp e}_y)$ and $\xi$ is the elliptic angle.  The inter-molecule potential can be written as\cite{Shi3, Xu} 
\begin{eqnarray}
V({\cp r})&=&\frac{C_3}{r^3} \Big(
	3\cos^2\theta -1 +3 f_{\xi}\sin^2\theta\cos(2\phi)\Big) \nonumber\\
	&&+ \frac{C_6}{r^6} \Big(1-[\cos^2\theta +f_{\xi} \sin^2\theta\cos(2\phi)]^2\Big). \label{V3d}
\end{eqnarray}
Here ${\cp r}=(r,\theta,\phi)$ is the relative distance; $f_{\xi}=\sin(2\xi)$; $C_3=\frac{d^2}{48\pi\epsilon_0(1+\delta_r^2)}$,  $C_6 = \frac{d^4}{128\pi^2\epsilon_0^2\Omega(1+\delta_r^2)^{3/2}}$ ($\delta_r = \frac{|\delta|}{\Omega}$) with $\Omega$ and $\delta$, respectively, denoting the frequency and detuning of microwave field; $d$ is the dipole moment. It has been shown that Eq.~(\ref{V3d}) can well approximate  realistic molecular interaction under the condition $\xi\le \pi/12$ and $\delta_r\ll 1$\cite{Shi3, Xu}, and therefore in this work we take a small fixed $\delta_r=0.2$.  In the limit of $\delta_r\rightarrow 0$, one can define an effective dipole moment  $d_{\rm eff}=d/\sqrt{12}$\cite{Zwierlein2022}, and similar to the definition in { these work\cite{Bohn5, Hutson4}}, $d_{\rm eff}$ gives rise to a dipole length $l_d\equiv \frac{m}{\hbar^2}\frac{d_{\rm eff}^2}{4\pi\epsilon_0}$, with $m$ the mass of single molecule. Specifically, we have $l_d=1.1,\ 2.6,\ 8.3,\ 1.3 \ (10^4a_0)$ for NaK\cite{Luo1,Luo2,Luo3}, NaRb\cite{Wang,Wang2}, NaCs\cite{Will1,Will2,Will_new}  and CaF\cite{Doyle} molecules, respectively. In comparison, the size of shielding core in these systems is much smaller ($\sim100-1000 a_0$ for $\Omega$ of tens of MHz), which determines the mean inter-molecule distance in a bound state. To facilitate numerical evaluations, in this work we take the length unit as $l_u=l_d/20$,  which is comparable to the typical size of bound state. Accordingly, the energy unit is $E_u=\frac{\hbar^2}{ml_u^2}$. Take NaK molecule as an example, we have $\hbar\Omega/E_u\approx157$ at $\Omega=(2\pi)30$MHz. 
We note that up to a constant factor, the length and energy units here are consistent with those in previous studies\cite{Bohn5, Hutson4}, under which the bound states can exhibit universal features\cite{Hutson4}.

\begin{figure*}[!t]
\centering
\includegraphics[width=\textwidth]{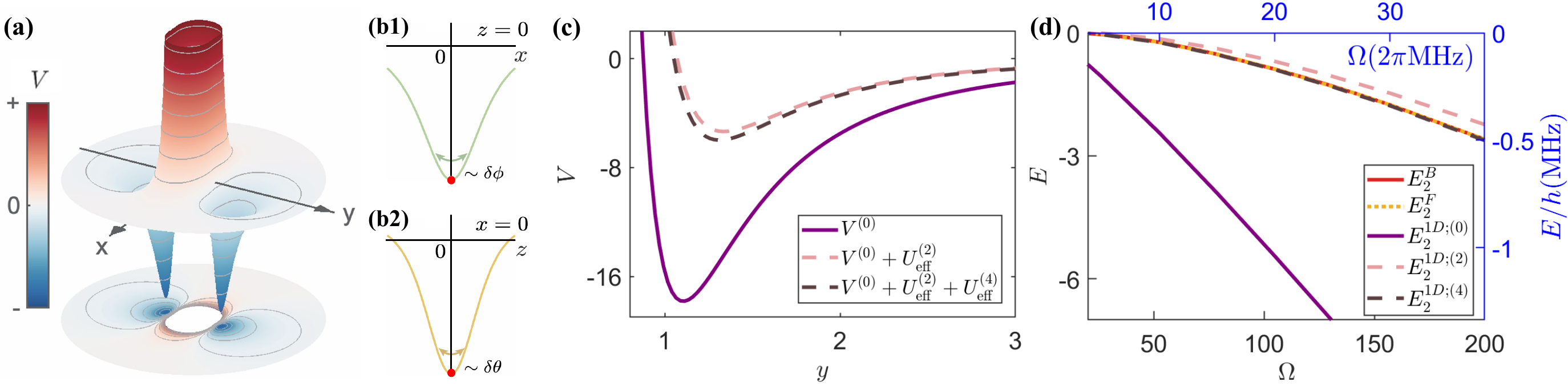}
\caption{{\bf{Interaction potential and tetratomic bound state of microwave-shielded molecules with elliptic angle $\xi=\pi/12$.} } (a) Typical interaction potential $V({\cp r})$ at $xy$ plane ($z=0$). (b1,b2) Schematics of angular fluctuations of $V({\cp r})$ along  $x$ ($\sim \delta\phi$) and $z$ ($\sim \delta\theta$) directions. The red points denote the minimum of $V$ when ${\cp r}$ locates at $y$ axis ($x=z=0$).  (c) Comparison of different 1D potentials (along $y$) at $\hbar\Omega/E_u=157$. $V^{(0)}$ is the bare potential, and $U_{\rm eff}^{(2)},\ U_{\rm eff}^{(4)}$ are the effective potentials from the lowest- and fourth-order angular fluctuations, respectively. (d) Tetratomic binding energies as functions of $\Omega$.  $E_{2}^{B}$ and $E_{2}^{F}$ are from exact numerical solutions of bosonic and fermionic molecules, and $E^{{\rm 1D};(n)}_2$ is from effective 1D model up to the $n$-th order angular fluctuations. The units of length, energy and $\Omega$ are respectively $l_u$, $E_u$ and $E_u/\hbar$. The upper and right axes in (d) show the absolute values of $\Omega$ and $E$ for NaK system.}  \label{fig_V}
\end{figure*}

Note that although the shielding core  ($\sim100-1000 a_0$) is much smaller than the dipole length ($l_d$), the core size determines the characteristic length scale of few-molecule bound states and therefore the core anisotropy cannot be neglected. This distinguishes our work from previous studies of few-body physics in dipolar systems\cite{Greene1,Greene2, Endo}, which assumed an isotropic s-wave potential at short range. 
Moreover, in contrast to our previous work on highly elliptic case\cite{SWC}, here we focus on small, finite ellipticities of microwave field, specifically  with $\xi\le \pi/12$. Such small $\xi$ not only validate $V(\cp r)$ as a good approximation for the real interaction potential\cite{Shi3, Xu}, but also produce a strong shielding effect that efficiently stabilizes the system against two-body losses\cite{Huston5}. In this regime, the repulsive $1/r^6$ shielding core exists in all directions, while  the long-range $-1/r^3$ attraction remains most pronounced along $y$, see Fig.~\ref{fig_V}(a). We will  establish effective 1D models along this direction to describe the bound state formation in this system.

{
\subsection*{Derivations of Effective 1D models} }

To construct an effective model along $y$, one must integrate out angular fluctuations in the perpendicular  directions. For smaller $\xi$,  the effect of angular fluctuations is stronger due to the weaker attraction of $V(\cp r)$ along $y$. In this case, it is necessary to include high-order angular fluctuations beyond the lowest-order terms considered previously\cite{SWC}. In the following, we derive these effective models with high-order angular fluctuations for tetratomic and hexatomic molecular systems.

{
\subsubsection*{Tetratomic system} }

In the basic two-molecule level with relative coordinate ${\cp r}=(r,\theta,\phi)$, we expand   
\begin{equation}
\theta=\theta^{(0)}+\delta\theta,\ \ \ \phi=\phi^{(0)}+\delta\phi, \label{angular_fluc}
\end{equation}
where $\theta^{(0)}=\pi/2,\ \phi^{(0)}=\pm \pi/2$ refer to  $\pm y$ direction, and  $\delta\theta, \delta\phi$ are angular fluctuations  beyond  this direction. 
As shown in Fig.~\ref{fig_V}(b1,b2), such angular fluctuations lead to the increase of $V(\cp r)$ along $z$ and $x$ directions. Here, we expand the interaction $V({\cp r})$ and kinetic term $H_k({\cp r})=-\hbar^2\nabla^2_{\cp r}/m$ up to the fourth-order of these fluctuation modes: $V=\sum_{n=0,2,4}V^{(n)}$ and $H_k=\sum_{n=0,2,4}H_k^{(n)}$, see Methods, where
$n=0$ represents bare terms that are independent of $\delta\theta$ and $\delta\phi$, and $n=2$ and $n=4$ respectively denote the second- and fourth-order corrections due to angular fluctuations. 

Under adiabatic representation\cite{Blume2,Cavagnero,Greene3}, the tetratomic wavefunction can be written as
\begin{equation}
\Psi_2({\cp r})=\frac{1}{r}\sum_{\nu} F_{\nu}(y) \psi_{\nu}(y;\delta\theta,\delta\phi), \label{psi2}
\end{equation}
with $y=r\sin\phi^{(0)}$ the reduced 1D coordinate. By integrating out the angular fluctuations in $\psi_{\nu}$, we can obtain an effective Hamiltonian governing the reduced 1D wavefunction $F_{\nu}$. Here we will only keep the ground state level ($\nu=0$) and neglect its coupling with higher levels ($\nu\ge 1$), since the bound states we consider are quite shallow as compared to the typical energy gap between different $\nu$, {see Supplementary Note 1 and Supplementary Figure 1}. Moreover, since different orders of fluctuations contribute differently to the effective potential, we adopt a separate treatment as below. First, we solve $\psi_{\nu=0}$ from 
\begin{equation}
\left( H_k^{(2)}+V^{(2)} \right) \psi_{0}(y;\delta\theta,\delta\phi) = U_{\rm eff}^{(2)}(y) \psi_{0}(y;\delta\theta,\delta\phi). \label{HO}
\end{equation}
The Hamiltonian here  ($H_k^{(2)}+V^{(2)}$) describes two independent harmonic oscillators in terms of $\delta {\theta}$ and $\delta\phi$, as shown in Methods. In this sense, these fluctuation modes are similar to the vibrational normal modes in polyatomic molecules\cite{normal_mode_1,normal_mode_2}, while the main difference lies in  that the emergence of our modes is due to anisotropic molecule-molecule interactions rather than the motion of atoms within a single molecule. 

Eq.(\ref{HO}) leads to a zero-point energy for the ground state $\nu=0$: (denoting $r\equiv |y|$)
\begin{equation}
U_{\rm eff}^{(2)}(r)=\left( \sqrt{1+f_{\xi}} + \sqrt{2 f_{\xi}}
\right) \sqrt{\frac{\hbar^2}{m} \left( \frac{3C_3}{r^5} + \frac{2 f_{\xi} C_6}{r^8} \right)}.  \label{Ueff2}
\end{equation}
This is the lowest-order correction to  1D potential, which  scales as $U_{\rm eff}^{(2)}\sim 1/r^4$ at $r\rightarrow 0$,  as previously also derived for $\xi=\pi/4$\cite{SWC}. 
In this work, we take one step further to evaluate the high-order correction as  
\begin{equation}
U_{\rm eff}^{(4)}(r)= \langle \psi_{0}| H_k^{(4)} +V^{(4)} |\psi_{0}\rangle. \label{U4}
\end{equation}
An analytical expression of $U_{\rm eff}^{(4)}$ is presented in Methods. At $r\rightarrow 0$,  $|U_{\rm eff}^{(4)}|\sim 1/r^2\ll 1/r^4 \sim U_{\rm eff}^{(2)}$; namely,  the high-order correction of effective potential is much smaller than the lowest-order one. This, in turn, validates the separate treatment of $U_{\rm eff}^{(n)}$ as above. 

Finally, we obtain an effective 1D Hamiltonian for tetratomic system: 
\begin{equation}
\left( - \frac{\hbar^2}{m} \frac{\partial^2}{\partial y^2} + V^{(0)}+U_{\rm eff}^{(2)} + U_{\rm eff}^{(4)} \right) F_{0}(y) = E_{2}^{\rm 1D}F_{0}(y). \label{Heff_2}
\end{equation}
Here $V^{(0)}$ is the bare interaction along $y$, and $U_{\rm eff}^{(2)},\ U_{\rm eff}^{(4)}$ are respectively the second- and fourth-order corrections from angular fluctuations. A typical comparison of interaction potentials up to different orders of corrections is shown in Fig.~\ref{fig_V}(c).  

{
\subsubsection*{Hexatomic system} }

For a hexatomic system of three molecules, the angular fluctuations of three tetratomic pairs can entangle with one another. Consequently,    the effective potential of hexatomic system generally cannot be formulated as a direct summation of three independent tetratomic potentials. 
Nevertheless, as we will see in this section, the lowest-order angular fluctuations of hexatomic system  always take the form of harmonic oscillators, while the high-order fluctuations can be evaluated within the lowest oscillator mode, similar to the tetratomic case. Specifically, here we will derive the effective 1D potential for two hexatomic systems. The first is a heavy-heavy-light system, where we focus on the induced heavy-heavy potential mediated by the light particle in the Born-Oppenheimer (BO) limit. This exactly solvable case serves as a benchmark for justifying the validity of effective 1D approach to hexatomic problem, and meanwhile, { it suggests the suppression of Efimov physics in this system}, as previously discussed for highly elliptic case\cite{SWC}.  We then proceed with the second system of three identical molecules. 

In the BO framework, we consider two localized heavy molecules at $\pm{\cp R}/2$, which interact with a moving light molecule via the microwave-shielded potential. The   Hamiltonian of the light object is   
\begin{equation}
H_L({\cp r})=-\frac{\hbar^2}{2m}\nabla^2_{\cp r} + V({\cp r}-\frac{\cp R}{2})+V({\cp r}+\frac{\cp R}{2}),\label{H_BO}
\end{equation}
giving the Schr{\"o}dinger equation 
$H_L({\cp r})\Psi_L({\cp r})=V_{\rm BO}({\cp R})\Psi_L({\cp r})$, where  the eigen-energy $V_{\rm BO}({\cp R})$ can be viewed as an effective potential between two heavy molecules due to the moving light object. Note that here we have neglected the diagonal BO correction due to the motion of heavy molecules, which is much smaller than $V_{\rm BO}$ given $M\gg m$, {see Supplementary Note 1 and Supplementary Figure 2}. Considering the interaction anisotropy as in Eq.~(\ref{V3d}),  we will focus on the case when ${\cp R}$ is along $y$ direction such that the bound state formation is most favored.

Similar to Eq.~(\ref{angular_fluc}), here we define two angular fluctuation modes, $\delta\theta$ and $\delta\phi$, for the light molecule with coordinate ${\cp r}=(r,\theta,\phi)$. Then $H_L$  can be expanded up to different orders of these modes, see Methods for $H_L^{(n)}$ up to $n=2$. The wavefunction of light molecule, $\Psi_L({\cp r})$, can be expanded as the form of Eq.~(\ref{psi2}). 
Again we only focus on the ground state level ($\nu=0$) and require $\psi_{\nu=0}$ follow 
\begin{equation}
H_L^{(2)} \psi_{0}(y;\delta\theta,\delta\phi) = U_{\rm L;eff}^{(2)}(y) \psi_{0}(y;\delta\theta,\delta\phi),
\end{equation}
where $y$ is the reduced 1D coordinate. Since $H_L^{(2)}$ is composed by two independent harmonic oscillators in terms of $\delta\theta$ and $\delta\phi$, we can analytically obtain the zero-point energy $U_{\rm L;eff}^{(2)}$, see Methods, which gives the effective heavy-heavy interaction  up to the lowest-order angular fluctuations. For $\xi=\pi/4$, it recovers the results obtained previously\cite{SWC}.

The evaluation of  high-order angular fluctuations, however, is quite complicated due to very complex form of  high-order expansion term ($H_{L}^{(4)}$). Considering  $H_{L}^{(4)}$ only contributes to a smaller correction as compared to $H_{L}^{(2)}$, we simplify its evaluation by neglecting angular entanglements between different heavy-light pairs. In this way,  the fourth-order contribution can be evaluated as 
\begin{equation}
U_{\rm L;eff}^{(4)}(y)=  u_{\rm eff}^{(4)}(r_{+})+ u_{\rm eff}^{(4)}(r_{-}),
\end{equation} 
where $u_{\rm eff}^{(4)}$ is the fourth-order correction of a heavy-light pair with relative distance $r_{\pm}=|y\pm R/2|$, see Methods. Finally, the reduced 1D equation for the light molecule can be written as
\begin{eqnarray}
&&\left( - \frac{\hbar^2}{2m} \frac{\partial^2}{\partial y^2}
 + V^{(0)}(r_{+})+V^{(0)}(r_{-})\right. \nonumber\\
&&\qquad\left. + U_{\rm L;eff}^{(2)} + U_{\rm L;eff}^{(4)}\right) F_{0}(y)
 = E_{L}^{\rm 1D}F_{0}(y). \label{Heff_L}
\end{eqnarray}
Here, the eigen-energy $E_{L}^{\rm 1D}$ is exactly the light-induced potential $V_{BO}$ at a given heavy-heavy separation $R$. 

For a hexatomic system of three identical molecules with coordinates $\{{\cp r}_1,{\cp r}_2,{\cp r}_3\}$, we can define two relative coordinates ${\cp r}={\cp r}_2-{\cp r}_1$ and $\bm{\rho}=\frac{2}{\sqrt{3}}({\cp r}_3-({\cp r}_1+{\cp r}_2)/2)$. The Hamiltonian in center-of-mass frame can be written as 
\begin{eqnarray}
H_{3}&=&-\frac{\hbar^2}{m} (\nabla^2_{\cp r}+\nabla^2_{\bm{\rho}})+V({\cp r})+V({\cp r}_+)+ V({\cp r}_-), \label{H3}
\end{eqnarray}
with 
${\cp r}_{\pm}=\frac{\cp r}{2}\pm\frac{\sqrt{3}{\bm{\rho}}}{2}$. Denoting the angular fluctuations of ${\cp r},\bm{\rho}$ as $\{\delta\theta_{r},\delta\phi_{r}\}$ and $\{\delta\theta_{\rho},\delta\phi_{\rho}\}$, one can then expand $H_3$ in terms of these fluctuation modes. The second-order expansion, $H_{3}^{(2)}$, is composed by two sets of independent harmonic oscillators in terms of $\{\delta\theta_r, \delta\theta_{\rho}\}$ and $\{\delta\phi_r, \delta\phi_{\rho}\}$,  see Methods. 

For three molecules  with a given projection ordering along $y$, their wavefunction can be written as
\begin{equation}
\Psi_3({\cp r},\bm{\rho})=\frac{1}{r\rho}\sum_{\nu} F_{\nu}(y_r,y_{\rho}) \psi_{\nu}(y_r, y_{\rho}; \delta\theta_r,\delta\phi_r,\delta\theta_{\rho},\delta\phi_{\rho}). \label{psi3}
\end{equation}
Here $y_r$ and $ y_{\rho}$ are the projections of ${\cp r}$ and $\bm{\rho}$ along $y$ direction. Up to the lowest-order fluctuations, the ground state ($\psi_{\nu=0}$) of the angular part satisfies $H_3^{(2)} \psi_{0} = U_{\rm 3;eff}^{(2)}(y_r,y_{\rho}) \psi_{0}$,
with $U_{\rm 3;eff}^{(2)}$ the effective 1D potential induced by the lowest-order angular fluctuations, as previously studied for $\xi=\pi/4$\cite{SWC}.  

Following the strategy in evaluating high-order correction to heavy-heavy-light system, here we neglect the entanglement between different tetratomic pairs and write the  fourth-order correction as 
\begin{equation}
U_{\rm 3;eff}^{(4)}(y_r,y_{\rho})=  U_{\rm eff}^{(4)}(r)+U_{\rm eff}^{(4)}(r_{+})+ U_{\rm eff}^{(4)}(r_{-}),
\end{equation} 
where $r=|y_r|$, $r_{\pm}=|y_r\pm \sqrt{3}y_{\rho}|/2$ and $U_{\rm eff}^{(4)}$ follows Eq.~(\ref{U4}). Finally, the effective 1D equation for three identical molecules is written as 
\begin{eqnarray}
&&\left[ - \frac{\hbar^2}{m}
 \big(\frac{\partial^2}{\partial y_r^2}+\frac{\partial^2}{\partial y_{\rho}^2}\big)
 + V^{(0)}(r)\right. \nonumber\\
&&\qquad\left. +V^{(0)}(r_{+})+V^{(0)}(r_{-})
 + U_{\rm 3;eff}^{(2)} + U_{\rm 3;eff}^{(4)}\right]F_{0}(y_r,y_{\rho}) \nonumber\\
&&\qquad= E_{3}^{\rm 1D}F_{0}(y_r,y_{\rho}). \label{Heff_3}
\end{eqnarray}
Here $U_{\rm 3;eff}^{(2)}$ and $U_{\rm 3;eff}^{(4)}$ are the effective 1D potentials from the lowest- and high-order angular fluctuations, which, respectively, scale as $1/r^4$ and $-1/r^2$ whenever pairwise molecules come close with distance $r\rightarrow 0$. Their explicit expressions are presented in Methods. 

\bigskip

{
\subsection*{Numerical results}

\subsubsection*{Tetratomic bound state} }

We have obtained the tetratomic bound states using both exact numerical solutions  and the effective 1D model, and from the comparison of these results, we are able to  identify the parameter regime for the validity of 1D description. To numerically solve the bound states, we have  expanded the tetratomic wavefunction in terms of eigenstates in momentum ($k$) and angular momentum ($l,m$) space and diagonalize the Hamiltonian using $\{klm\}$ basis\cite{SWC}. Meanwhile, we  have diagonalized the effective 1D model in Eq.~(\ref{Heff_2}) directly in real space to obtain the bound state solution.

Take a small ellipticity $\xi=\pi/12$ as an example, in Fig.~\ref{fig_V}(d) we show the binding energies of the tetratomic ground state from exact solutions and from effective 1D models up to different orders of angular fluctuations. The corresponding 1D potentials along $y$ are shown in Fig.~\ref{fig_V}(c). We can see that the bare potential ($V^{(0)}$) significantly overestimates the binding energy, while the effective potential with lowest-order correction ($V^{(0)}+U_{\rm eff}^{(2)}$) yields a substantial improvement. Encouragingly, by further incorporating the fourth-order correction ($U_{\rm eff}^{(4)}$), the effective 1D model achieves excellent agreement with exact results across a wide range of $\Omega$. We emphasize that including high-order corrections is more crucial for smaller $\xi$, where the interaction anisotropy is less pronounced and the fluctuations beyond $y$ direction can affect the bound state formation more significantly. 

Despite accurately predicting the tetratomic state in Fig.~\ref{fig_V}(d), the  effective 1D model does not always perform well, particularly for shallow bound states with small binding energies\cite{SWC}.  In the following, we identify the validity regime of effective 1D model by comparing with exact solutions of tetratomic system. Before proceeding, we note that  the hard-core nature of 1D model in Eq.~(\ref{Heff_2}) enables an important eigen-state property as Bose-Fermi duality, i.e., bosonic and fermionic molecules share identical eigen-spectrum and real-space distribution. This occurs because, for a given spatial ordering of particles on a 1D line, the system wavefunction is independent of statistics, with different statistics only determining the sign change of the wavefunction in other spatial ordering sectors.  Such duality has been previously pointed out in short-range interacting 1D  systems\cite{Girardeau1,Cheon} and  long-range interacting dipolar systems\cite{Endo,SWC}. Here, we utilize Bose-Fermi duality as a criterion for judging the validity of 1D model.  

\begin{figure}[!tbp]
\centering
\includegraphics[width=7cm]{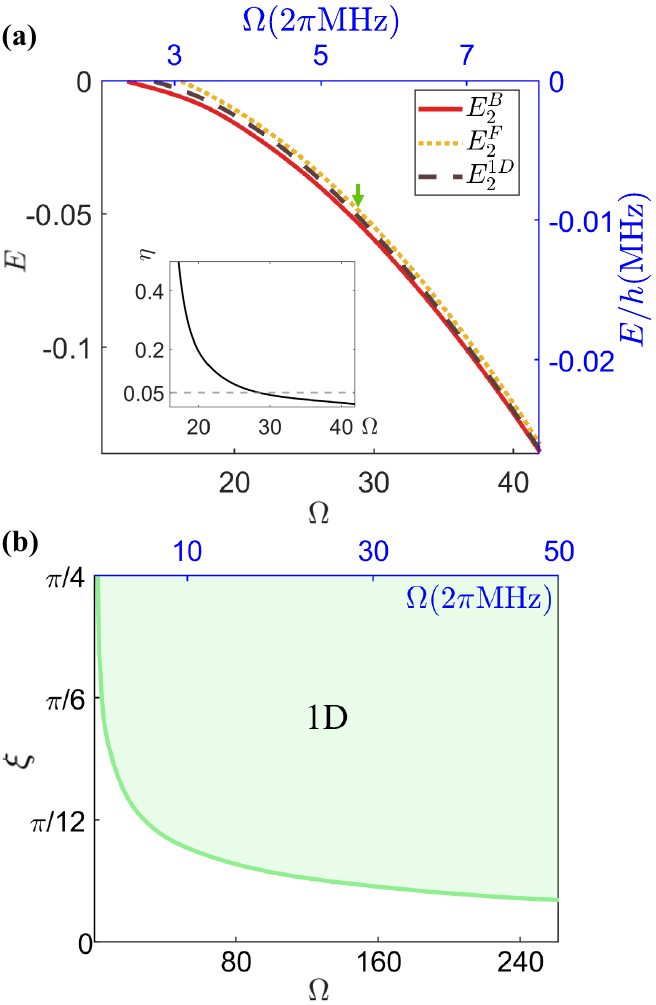}
\caption{{\bf{Validity of an effective 1D description. } }(a) Tetratomic binding energy as a function of $\Omega$ at a given $\xi=\pi/12$. $E_{2}^B$ and $E_{2}^F$ are exact solutions of bosonic and fermionic systems, and $E_{2}^{\rm 1D}$ is from effective 1D model (\ref{Heff_2}) with high-order angular fluctuations. Inset plot shows $\eta$, defined in (\ref{eta}), as a function of $\Omega$. The horizontal line marks the location when $\eta$ reduces to $5\%$, as also shown by the green arrow in the main plot. The upper and right axes  show the absolute values of $\Omega$ and $E_2$ for NaK system. 
(b) Diagram for the validity of 1D description in $(\Omega,\xi)$ plane. Solid line shows the critical boundary when $\eta=5\%$, and the 1D description is a good approximation in the green area above this boundary. The units of energy and $\Omega$ are respectively $E_u$ and $E_u/\hbar$. The upper axis of $\Omega$ in (b) shows its absolute value for NaK system.  }  \label{fig_validity}
\end{figure}

In Fig.~\ref{fig_validity}(a), we show a magnified comparison of tetratomic binding energies between exact results  and  1D predictions based on Eq.~(\ref{Heff_2}), for a fixed $\xi=\pi/12$ and varying $\Omega$. We can see that near the threshold of bound state formation, the 1D model  fails for both bosonic  and fermionic  systems, as evidenced by visible deviations between the exact ($E_{2}^{B}$, $E_{2}^{F}$) and 1D ($E_{2}^{1D}$) results. Physically, this failure arises  because these shallow bound states exhibit a 3D character. For instance, the emergence of $E_{2}^{B}$ for bosonic molecules is associated with an s-wave resonance where the scattering length diverges,  $a_s\rightarrow\infty$.  As explored in detail for $\xi=\pi/4$\cite{SWC}, the shallow bound states near resonance  well follow $E_{2}^{B}= -\hbar^2/(ma_s^2)$, and thus these states expand isotropically in free space rather than being confined essentially to one direction. For fermionic molecules, the shallow tetratomic bound state resides  in the p-wave channel, which emerges at a larger $\Omega$ than the bosonic one. However, as $\Omega$ increases, both bosonic and fermionic systems gradually depart from the 3D regime and begin to exhibit Bose-Fermi duality, consistent with the 1D description. As shown in Fig.~\ref{fig_validity}(a), as $\Omega$ increases, $E_{2}^{B}$ and $E_{2}^{F}$ converge and tend to merge into a single line predicted by $E_{2}^{1D}$.  To quantify this convergence, we define a dimensionless parameter 
\begin{equation}
\eta=\left|\frac{E_{2}^{B}-E_{2}^{F}}{E_{2}^{B}+E_{2}^{F}}\right|, \label{eta}
\end{equation}
which represents the relative deviation between $E_{2}^{B}$ and $E_{2}^{F}$. As shown in the inset of Fig.~\ref{fig_validity}(a), $\eta$ decreases rapidly with increasing $\Omega$.  We take the location where $\eta$  reduces to $5\%$ as the critical boundary for the validity of 1D description. In Fig.~\ref{fig_validity}(b), we map out this boundary in the parameter plane of $(\Omega,\xi)$. The green area above this boundary is the validity regime of 1D model, where $\eta\le 5\%$. Then, we can see that the 1D model is valid across a wide range of $\Omega$ even at small $\xi$. Take NaK system as an example,  the validity just requires $\Omega>(2\pi)5.4$MHz for $\xi=\pi/12$, and $>(2\pi)38.3$MHz for $\xi$ as small as $\pi/30$. This shows that in practical situations with $\Omega$ of tens of MHz,  the 1D model performs  robustly well even for small microwave field ellipticities. 
In comparison, for NaRb and NaCs systems the same $\Omega/E_u$ will generate much smaller $\Omega$, and thus for the same $\xi$ these two systems will enter 1D regime at much smaller $\Omega$, {see Supplementary Note 2 and Supplementary Figure 3}.  Importantly, the bound states within the 1D region can be quite shallow, which therefore are not expected to cause severe loss in realistic experiments. For instance, at the 1D boundary in Fig.~\ref{fig_validity}(b), we have $|E_2^{1D}|/h\le 50 $KHz for NaK system at $\Omega\le (2\pi)50$MHz, {see Supplementary Figure 4}, which is associated with very low decay rate to be well probed in experiment\cite{Luo3}. The tetramer binding energies at the 1D boundary for different molecular systems can be found in {Supplementary Figure 4}.  
Surely for those bound states far from the 1D boundary, they could be sufficiently deep with high decay rate and thus difficult to probe in practice. 

\begin{figure}[!tbp]
\centering
\includegraphics[width=8cm]{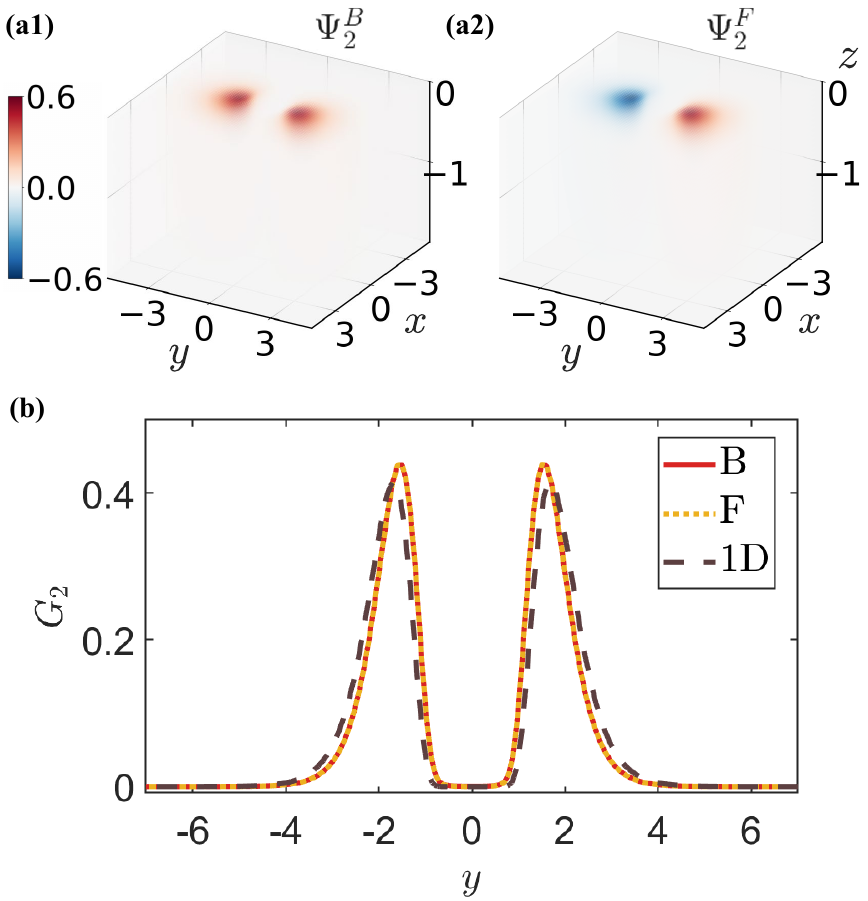}
\caption{{\bf{Bose-Fermi duality of tetratomic bound states.}} (a1,a2) show the real-space wavefunctions of bosonic ($\Psi_2^{B}$) and fermionic ($\Psi_2^{F}$) systems from exact solutions, and (b) shows their reduced density correlation functions $G_2(y)$ (Eq.\ref{G2}), in comparison with results from the effective 1D model. Here we take $\xi=\pi/12, \hbar\Omega/E_u=157$, and the length unit as $l_u$. 
}  \label{fig_duality}
\end{figure}

Similar to the highly elliptic case\cite{SWC}, the Bose-Fermi duality can be probed via the density-density correlation function along $y$:
 \begin{equation}
G_2(y)\equiv \langle n(0)n(y)\rangle. \label{G2}
\end{equation}
As seen from its definition, $G_2(y)$ directly measures the probability  of finding two molecules with relative distance $y$ along the 1D direction. For the tetratomic system, we have $G_2(y)=\int dxdz |\Psi_2({\cp r})|^2$ from exact 3D solutions and $=|F_{0}(y)|^2$ from effective 1D model. Taking a small $\xi=\pi/12$ and a typical $\hbar\Omega/E_u= 157$, in  Fig.~\ref{fig_duality}(a,b) we plot  the real-space wavefunctions of bosonic and fermionic systems, with their corresponding $G_2(y)$ shown in  Fig.~\ref{fig_duality}(c).  We can see that the exact $G_2$ for both bosonic and fermionic systems match very well with the 1D prediction,  directly demonstrating Bose-Fermi duality in this case. 

\begin{figure}[!tbp]
\centering
\includegraphics[width=8.5cm]{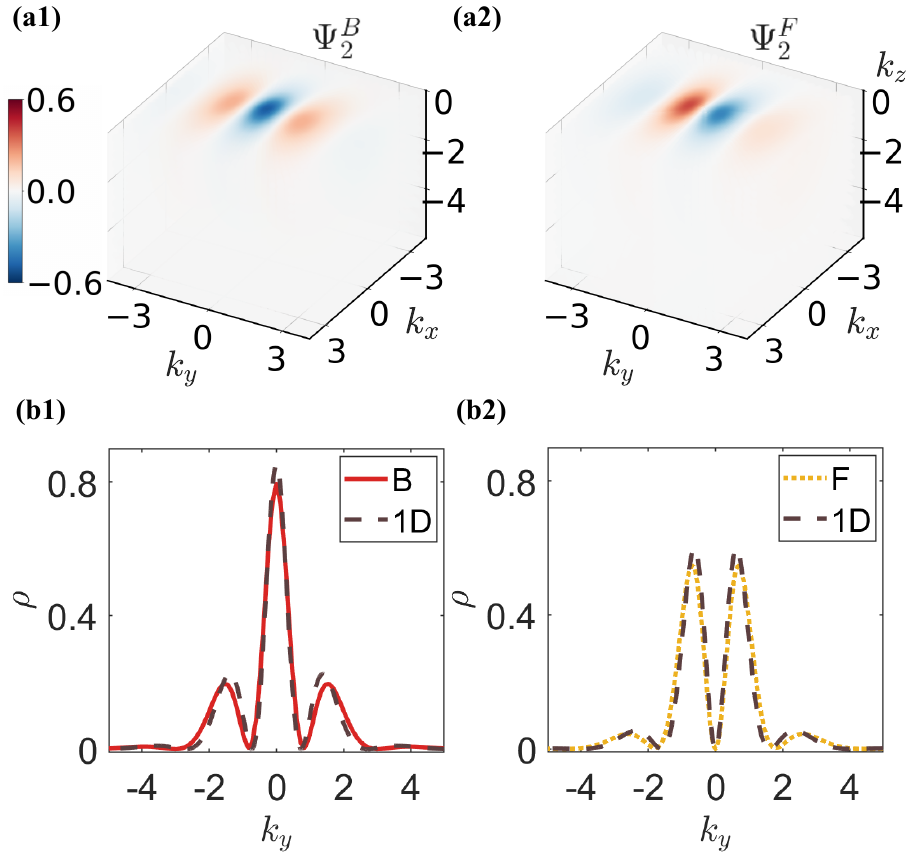}
\caption{{\bf{Momentum-space distributions of tetratomic bound states.}}
(a1,a2) show momentum-space wavefunctions  for the lowest tetratomic bound states in bosonic ($\Psi_2^{B}$) and fermionic ($\Psi_2^{B}$) systems at $\xi=\pi/12$ and $\hbar\Omega/E_u=157$. Accordingly, (b1,b2) show the reduced momentum-space distributions along $k_y$, from both exact solutions and effective 1D model. The length unit is $l_u$.
}  \label{fig_k}
\end{figure}

Despite duality in spectral and real space, bosonic and fermionic systems are typically distinguishable in momentum space. This is because the momentum-space wavefunction involves a transformation of real-space counterpart across different spatial ordering sectors, thereby successfully carrying  the information of exchange statistics.  As shown in Fig.~\ref{fig_k} (a1,a2), the two tetratomic systems indeed exhibit distinct momentum-space wavefunctions $\Psi_2^B({\cp k})$ and $\Psi_2^F({\cp k})$. Specifically, $\Psi_2^B$ is most pronounced at ${\cp k}=0$ while $\Psi_2^F$ is peaked at finite ${\cp k}$. As a result, the reduced momentum distributions along $y$, denoted by $\rho^{B/F}(k_y)=\int dk_xdk_z |\Psi_2^{B/F}({\cp k})|^2$, are also very different, see Fig.~\ref{fig_k}(b1,b2). Again,  the exact $\rho^{B/F}$ agree very well with the predictions from effective 1D theory. 

The real-space density correlation may be probed via  quantum gas microscopes as implemented in ultracold atoms\cite{correlation_ENS, correlation_MIT1, correlation_MIT2}, where the atoms are suddenly pinned on a deep optical lattice to enable single-atom imaging. The momentum distribution can be measured through time-of-flight imaging after dissociating the bound states, as carried out already for tetratomic molecules in NaK system\cite{Luo3}. 

{
\subsubsection*{Born-Oppenheimer potential} }

In the {subsection "Tetratomic bound state"}, we have examined the validity of effective 1D model for  tetratomic bound state. This validity, however, does not automatically  guarantee equal validity for hexatomic system. 
As we learn from few-body physics in atomic gases, three-body systems can behave qualitatively differently from two-body ones. For example, three particles can support Efimov physics as facilitated by a scale-invariant $-1/R^2$ potential, which is absent in two-body sector.  Therefore, it is necessary to independently examine the validity of 1D description for hexatomic system. In our previous study, we utilized the heavy-heavy-light molecular system to examine the Efimov physics and confirm the validity of effective 1D approach for a high microwave field ellipticity ($\xi=\pi/4$)\cite{SWC}. Here we evaluate the validity of 1D approach to  more realistic case of small $\xi$. 

\begin{figure}[!tbp]
\centering
\includegraphics[width=8cm]{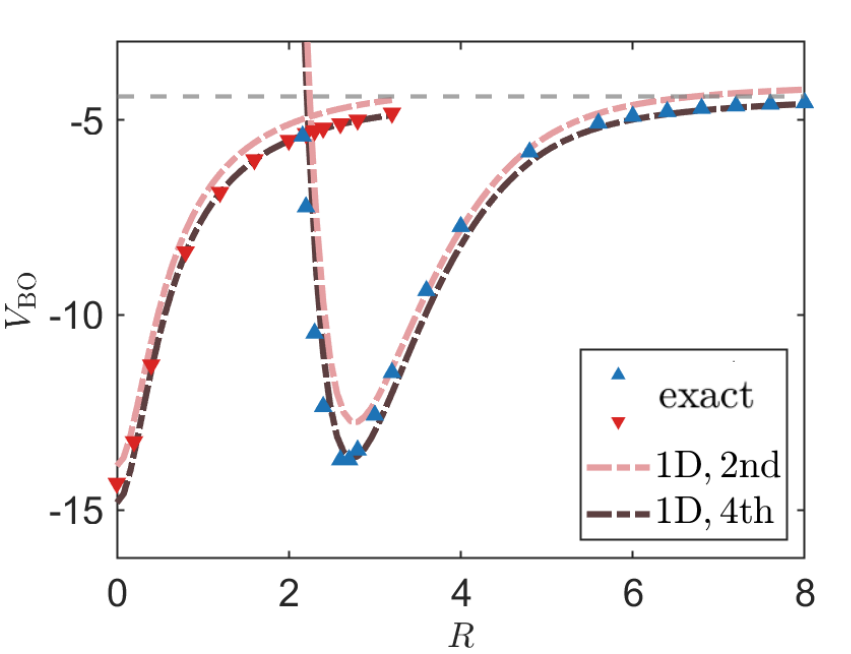}
\caption{{\bf{Born-Oppenheimer potential $V_{\rm BO}({\cp R}=R\hat{y})$ between two heavy molecules mediated by the light one.}} Here we take $\xi=\pi/12$ and $\hbar\Omega/E_u=157$. Red and blue triangles show two orthogonal eigen-levels of $V_{\rm BO}$ by solving the Schr{\"o}dinger equation of a light molecule. These levels correspond to the light molecule residing besides and in-between heavy ones, respectively.  
Pink and purple dashed lines are predictions from effective 1D models up to the second- and fourth-order angular fluctuations. The horizontal gray line shows the tetratomic binding energy of a heavy-light pair. The length and energy units are respectively $l_u$ and $E_u$.}  \label{fig_BO}
\end{figure}

Taking a small $\xi=\pi/12$ and a typical $\hbar\Omega/E_u=157$, we have numerically solved  the Schr{\"o}dinger equation of a light molecule to obtain the light-mediated heavy-heavy potential $V_{\rm BO}({\cp R})$, with ${\cp R}=R\hat{y}$ pointing along the most attractive ($y$) direction. The two lowest levels of $V_{\rm BO}$ are shown in  Fig.~\ref{fig_BO} as functions of $R$, see red and blue triangles. As discussed in Ref.~\cite{SWC}, these two orthogonal levels correspond to the light molecule residing besides (red) and in-between (blue) heavy ones, respectively.  
In Fig.~\ref{fig_BO}, these  results are compared with predictions from effective 1D models up to the lowest- and high-order angular fluctuations. We can see that the inclusion of high-order fluctuations further improve the accuracy of 1D prediction compared to the lowest-order case, leading to  excellent agreements with numerical results across the entire regime of $R$.

We have also checked other parameters of $\xi$ and $\Omega$, and found the 1D model with high-order angular fluctuations accurately predicts $V_{\rm BO}$ of heavy-heavy-light system for $\xi$ and $\Omega$ lying within the green area of Fig.~\ref{fig_validity}(b). In this regime, the bound states are sufficiently deep and the system is far from scattering resonance with $a_s\rightarrow\infty$. This is why the Efimov physics is greatly suppressed while universal bound states, which depend only on the physical parameters $l_d, \ \xi$ and $\Omega$, are favored. The formation of these bound states is uniquely facilitated by the attractive potential along $y$,  rather than by the large positive $a_s$ near resonance. 

{
\subsubsection*{Hexatomic bound state of three identical molecules} }

Given the accuracy of effective 1D model in reproducing tetratomic bound state and BO potential of heavy-heavy-light system, we can now confidently utilize it to study hexatomic bound state of three identical molecules. Here we take the parameters of $\xi$ and $\Omega$ within the validity regime of 1D description (the green area in Fig.~\ref{fig_validity}(b)) and numerically diagonalize the effective 1D Hamiltonian (Eq.~(\ref{Heff_3})) in discretized real space to obtain the hexatomic eigenstates.  Given the hard-core nature of 1D potential in Eq.~(\ref{Heff_3}), the bosonic and fermionic systems obey Bose-Fermi duality, possessing  identical energy spectra and real-space density profiles.

\begin{figure}[!tbp]
\centering
\includegraphics[width=8cm]{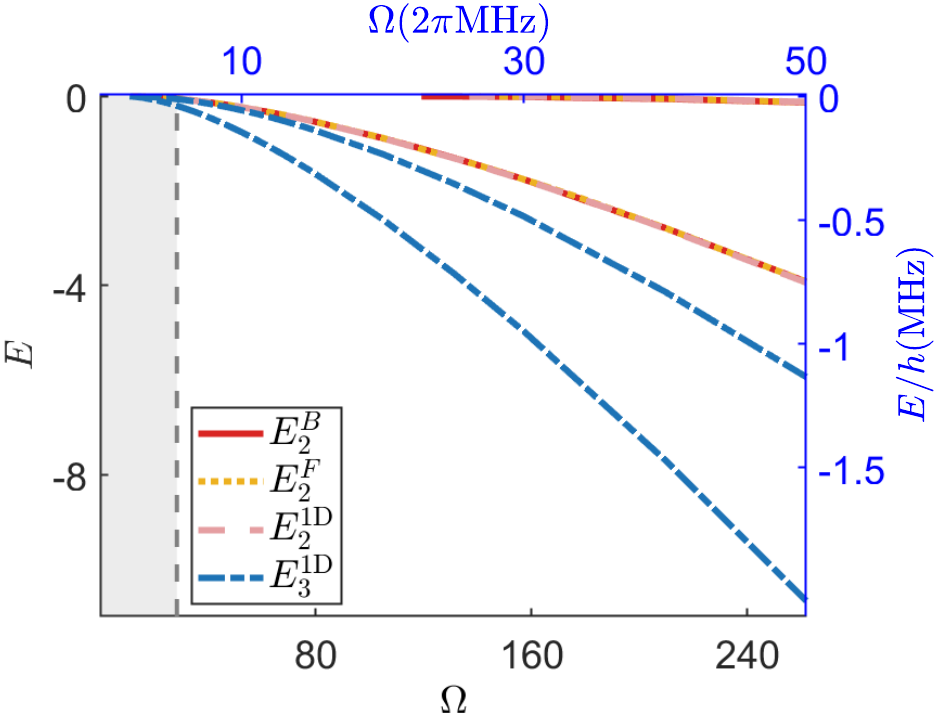}
\caption{{\bf{Tetratomic and hexatomic bound state spectra at $\xi=\pi/12$.}} 
 $E_{2}^B$ and $E_{2}^F$ are the exact tetratomic energies for bosonic and fermionic systems, in comparison with $E_{2}^{1D}$ from effective 1D model (\ref{Heff_2}). $E_{3}^{1D}$ is the hexatomic energy from effective 1D model (\ref{Heff_3}).  For each of them, we have shown two lowest energy levels. The shaded area marks small $\Omega$ regime where the 1D description is invalid, i.e., lying in the white region of Fig.~\ref{fig_validity}(b). The units of $\Omega$ and $E$ are respectively $E_u/\hbar$ and  $E_u$. Upper and right axes show the absolute values of $\Omega$ and $E$ for NaK system.}  \label{fig_E}
\end{figure}

Fig.~\ref{fig_E} shows the bound state spectra of tetratomic and hexatomic systems as functions of $\Omega$ for a fixed $\xi=\pi/12$. Then, we see that the hexatomic binding energies are  generally much deeper than the tetratomic ones, for both ground and excited states. We emphasize that these hexatomic bound states are universal, not Efimov-type, as their binding energies do not depend on short-range details but only on the physical parameters of dipole length ($l_d$) and microwave parameters ($\Omega,\ \xi$).  Furthermore, they differ significantly from universal bound states in atomic systems, where a finite mass imbalance is required for trimer binding to dominate over dimer binding\cite{KM, Blume, Petrov,  Pricoupenko, Parish,Cui, KM_1D,Mehta_1D,Petrov_1D}. In contrast,  the universal hexatomic binding revealed here requires no mass imbalance, a unique consequence of long-range interaction as we will discuss later. 

\begin{figure*}[!t]
\centering
\includegraphics[width=17cm]{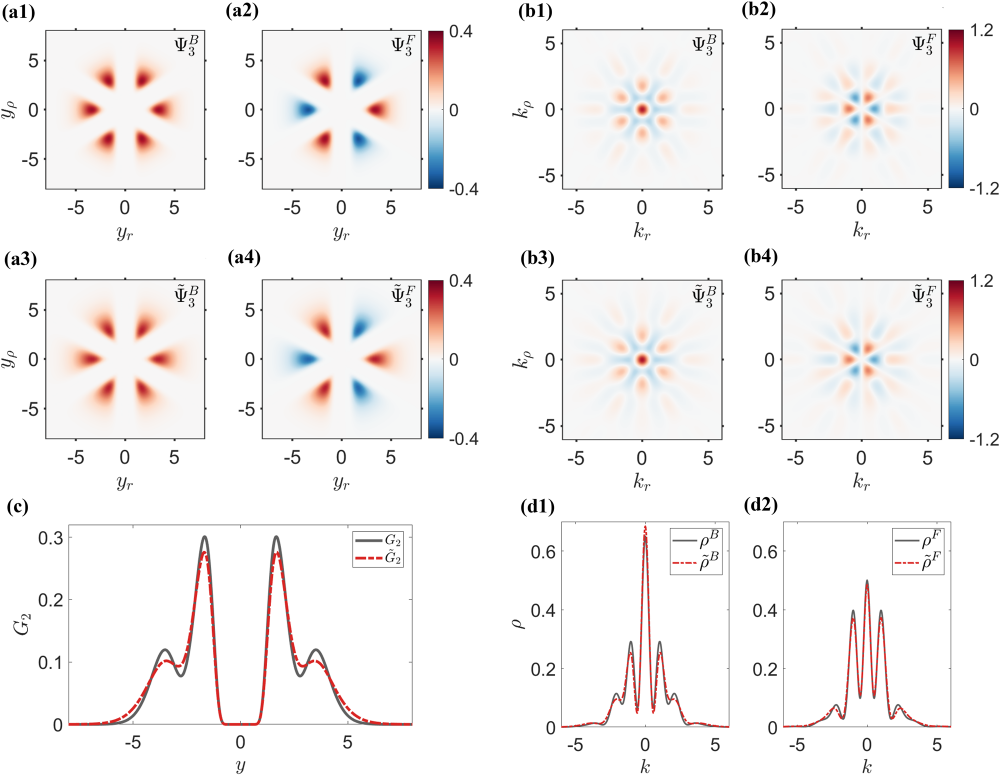}
\caption{{\bf{Real- and momentum-space distributions of hexatomic ground state.}} Here we take $\xi=\pi/12$ and $\hbar\Omega/E_u=157$.
(a1,a2) and (b1,b2) show real- and momentum-space wavefunctions  for bosonic ($\Psi_3^B$) and fermionic ($\Psi_3^F$) systems from the effective 1D model. In comparison, {(a3,a4) and (b3,b4)} show results from the simplified wavefunction in (\ref{simp}). (c) shows the density correlation function along $y$ from the 1D model ($G_2$) and from the simplified wavefunction ($\tilde{G}_2$).  (d1,d2) are the momentum distributions of bosonic ($\rho^B$)  and fermionic  ($\rho^F$) molecules from the 1D model, in comparison with results from the simplified wavefunction ($\tilde{\rho}^B$, $\tilde{\rho}^F$). The length unit is $l_u$.}  \label{fig_3body}
\end{figure*}

The effective 1D description ensures the Bose-Fermi duality for hexatomic molecules. As shown in Fig.~\ref{fig_3body}(a1,a2), the wavefunctions of bosonic ($\Psi_3^B$) and fermionic ($\Psi_3^F$) systems differ only by symmetry in real space, while their density correlation functions, $G_2(y)$, are exactly identical, see Fig.~\ref{fig_3body}(c). The different exchange symmetries lead to distinct momentum-space wavefunctions, as shown  in Fig.~\ref{fig_3body}(b1,b2) for bosonic and fermionic systems respectively. These distinctions can be probed via the single-particle momentum distribution along $y$, defined for a hexatomic system  as 
\begin{equation}
\rho^{B/F}(k)=\int  dk_{\rho} |\Psi_3^{B/F}(k_r=-k-k_{\rho}/\sqrt{3}, k_{\rho})|^2.
\end{equation}
As shown in Fig.~\ref{fig_3body}(d1,d2), $\rho^B$ and $\rho^F$ both display visible crystalline patterns, but they are still well distinguishable:  $\rho^B$ is more pronounced at $k_y=0$ while $\rho^F$ is more equally distributed among different peaks. 

Now we analyze why hexatomic binding is more favorable than tetratomic binding for microwave-shielded molecules.  
As previously discussed for $\xi=\pi/4$\cite{SWC}, the hexatomic ground state can be well approximated by the product of two tetratomic states linked side-by-side.  For three molecules aligned along $y$ in the order $y_1<y_2<y_3$, its ground state can be approximated as
\begin{equation}
\tilde{\Psi}_{3;g}=\Psi_{2;g}(y_{12})\Psi_{2;g}(y_{23}), \label{simp}
\end{equation}
where $y_{ij}\equiv y_i-y_j$ is the relative distance, and $\Psi_{2;g}(y)$ is the ground state of tetratomic molecules with relative distance $y$. We find that the relation in Eq.~(\ref{simp}) remains valid for small $\xi$. Taking $\xi=\pi/12$ and a typical $\hbar\Omega/E_u=157$,   we compare the hexatomic wavefunction $\Psi_{3;g}$ with the simplified form $\tilde{\Psi}_{3;g}$ in Fig.~\ref{fig_3body}(a1,a2,b1,b2) and {(a3,a4,b3,b4)} for both bosonic and fermionic systems in real- and momentum space, finding reasonably good agreement. The agreement is also evident from the density correlation function and momentum distribution plotted in Fig.~\ref{fig_3body}(c) and (d1,d2).    

Given the robust  relation in Eq.~(\ref{simp}), we can anticipate that the binding energy of the lowest hexatomic state is around twice the tetratomic energy, leading to a much deeper binding of hexatomic system as shown in Fig.~\ref{fig_E}. 
Physically, the universal phenomenon of enhanced hexatomic binding relative to  tetratomic one is a direct consequence of the long-range interaction potential.   
Moreover, the  rich crystalline patterns of hexatomic wavefunctions in both real- and momentum-space,  shown in Fig.~\ref{fig_3body}, are also uniquely associated with long-range interactions.  
From a few-to-many perspective, these patterns of few-molecule systems may shed light on intriguing crystalline orders in large molecular ensembles.

The extension of Eq.~(\ref{simp}) to large ensemble of molecules is straightforward: the ground state of $N$ molecules is a product of $N-1$ nearest-neighbor tetratomic states linked along $y$ direction, i.e., $\tilde{\Psi}_{N;g}=\Psi_{2;g}(y_{12})\Psi_{2;g}(y_{23})...\Psi_{2;g}(y_{N-1,N})$ for the spatial ordering $y_1<y_2...<y_N$. The mean inter-molecule distance in $\tilde{\Psi}_{N;g}$ is roughly given by the minimum of effective 1D potential in Fig.~\ref{fig_V}(c), which also determines the equilibrium density of this large bound state. In this sense,  $\tilde{\Psi}_{N;g}$  constitutes  a self-bound droplet with crystalline pattern along $y$ direction. Visually, this state can be called as a ``single-molecule array", which is universally applicable to both bosonic and fermionic systems. For fermionic system, this state is in  sharp contrast  to the conventional expectation of p-wave pairing superfluid as anticipated in previous experiment\cite{Luo3}. 

\begin{figure}[!tbp]
\centering
\includegraphics[width=8.5cm]{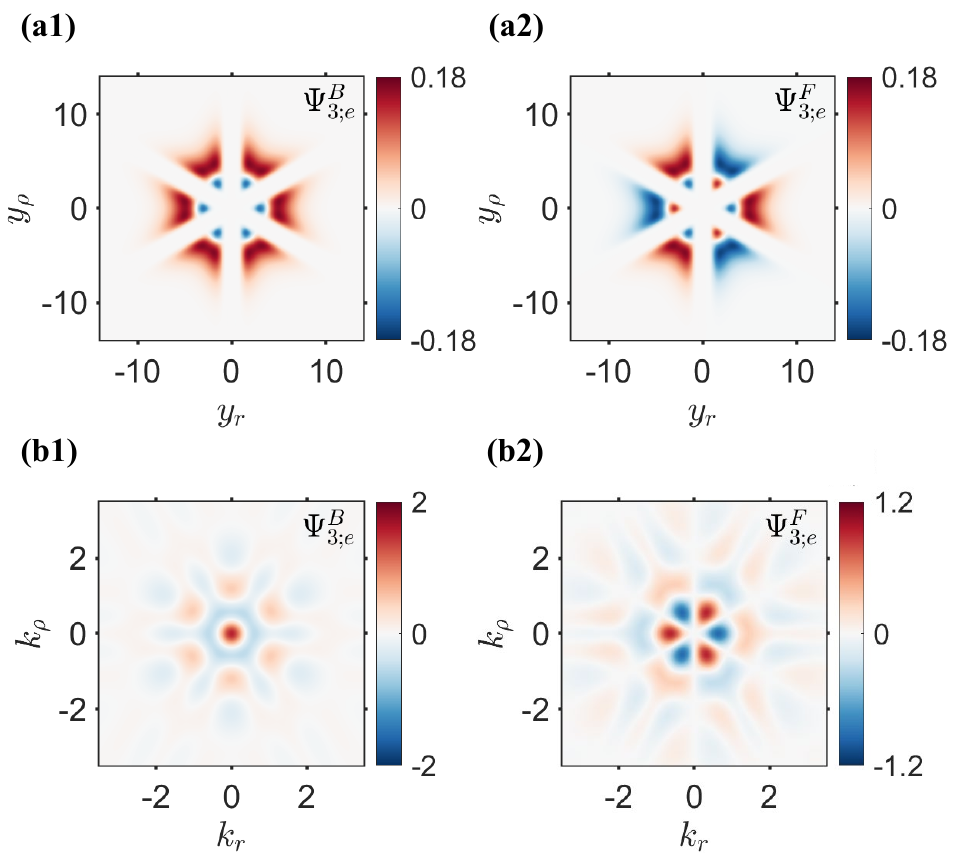}
\caption{{\bf{Real- and momentum-space wavefunctions of the first excited hexatomic state. }} Here we take $\xi=\pi/12$ and $\hbar\Omega/E_u=157$. (a1,a2) are the real-space wavefunctions of bosonic ($\Psi_{3;e}^B$) and fermionic ($\Psi_{3;e}^F$) systems, and (b1,b2) are their according momentum-space wavefunctions. The length unit is $l_u$.    
}  \label{fig_hex}
\end{figure}

We have also computed the first excited hexatomic bound state, with spectrum shown in Fig.~\ref{fig_E}. In Fig.~\ref{fig_hex}, we plot its typical real- and momentum-space wavefunctions for both bosonic and fermionic systems. Compared to  the ground state  in Fig.~\ref{fig_3body},  the excited state displays very different crystalline patterns with more nodes in  real and momentum space. In principle, this  hexatomic excited state could be related to the tetratomic excited state, i.e., by replacing  one tetratomic link from ground ($\Psi_{2;g}$) to excited state ($\Psi_{2;e}$). However, due to the very shallow $\Psi_{2;e}$ in this case (see Fig.~\ref{fig_E}), the excited branch can no longer be considered as an isolated level for constructing the hexatomic excited state, which should instead involve a large amount of contributions from low-lying scattering states. Therefore, we do not explicitly construct the hexatomic excited state from isolated tetratomic levels here.  

Finally, we discuss the possibility of observing Efimov trimers in bosonic molecules. As previously studied in dipolar systems\cite{Greene1}, Efimov trimers can be supported near an s-wave resonance under the condition $|a_s|\gg l_d$. However, for  current ultracold molecules with very large $l_d\sim 10^4a_0$\cite{Luo1,Luo2,Luo3, Wang, Will1,Will2, Doyle}, this condition is satisfied only within a narrow parameter region near resonance, as previously examined for $\xi=\pi/4$\cite{SWC}. In the present case with smaller $\xi$, such as $\xi=\pi/12$, the parameter region for $|a_s|\gg l_d$ is still quite narrow, which makes the exploration of Efimov physics in bosonic molecules extremely challenging. In comparison, the universal clusters revealed here are far more robust and practical to detect in realistic bosonic and fermionic molecular gases.  
{ On the other hand, we note that a very recent study has reported Efimov effect in bosonic molecules shielded by double microwave fields\cite{Greene_new}, where the $\sigma$ and $\pi$ lights produce a cancellation effect to the long-range  potential and a reasonably wide resonance region ($|a_s|\gg l_d$) can be found.   }

{
\section*{Conclusions} }

In summary, we have established a robust effective 1D theory for bound-state formation in 3D ultracold molecules under elliptic microwave shielding. Our results highlight the dimensional reduction purely through the anisotropy of long-range interaction potential, rather than through the application of any external confinement. A key finding is that high-order angular fluctuations have to be incorporated into the 1D model to achieve quantitative accuracy. This effective 1D description is robust across a wide range of coupling strengths at small ellipticities (Fig.~\ref{fig_validity}(b)). The 1D framework directly ensures a Bose-Fermi duality in tetratomic and hexatomic bound states, which can be directly probed  via spectroscopy or correlation measurements. We also  predict rich crystalline patterns in both real and momentum space, a unique feature supported by long-range interactions. Furthermore, by successfully constructing the hexatomic ground state from tetrameric units, we expect the ground state of a large system to be a self-bound droplet of many single molecules distributing along one direction with equal spacing. Such a ``single-molecule array" is universally applicable to both bosonic and fermionic molecular systems.

A natural extension of this work is to investigate few-molecule bound states in the limit of vanishing ellipticity ($\xi \to 0$). In particular, a key question is the evolution of bound states as $\xi$ is tuned to zero, a regime where the 1D description becomes inapplicable (see Fig.~\ref{fig_validity}(b)). For $\xi \rightarrow 0$, an effective 2D theory can be established by incorporating longitudinal fluctuations beyond the 2D plane, analogous to our present approach. Consequently, tuning $\xi$ can drive a dimensional crossover from 1D to 2D, potentially giving rise to distinct crystalline patterns and a breakdown of Bose-Fermi duality. Moreover, another extension of the present work is to consider the dual microwave fields as implemented in bosonic molecules\cite{Will2, Will_new,Wang2}. Following the few-to-many roadmap, a solid few-body study will offer a trustable path to explore complex many-body phases, a strategy particularly valuable for the field of ultracold molecules where few-body physics remains largely uncharted.

\bigskip
{
\section*{Methods} 
\subsection*{Tetratomic system} }

Up to the fourth-order of angular fluctuations, the interaction potential $V({\cp r})$ gives
\begin{eqnarray}
&&V^{(0)}(r)= -\big(1+3f_{\xi}\big)\frac{C_3}{r^3}
 +(1-f_{\xi}^2)\frac{C_6}{r^6}; \nonumber\\
&&V^{(2)}(r,\delta\theta,\delta\phi) \nonumber\\
&&\quad=\Big(\frac{3C_3}{r^3}+\frac{2f_{\xi}C_6}{r^6}\Big)
 \Big((1+f_{\xi})\delta\theta^2+2f_{\xi}\delta\phi^2\Big); \nonumber\\
&&V^{(4)}(r,\delta\theta,\delta\phi) \nonumber\\
&&\quad=-(1+f_{\xi})\Big(\frac{C_3}{r^3}
 +(1+\frac{5f_{\xi}}{3})\frac{C_6}{r^6}\Big)\delta\theta^4 \nonumber\\
&&\qquad-2f_{\xi}\Big(\frac{C_3}{r^3}+\frac{8f_{\xi}C_6}{3r^6}\Big)\delta\phi^4 \nonumber\\
&&\qquad-2f_{\xi}\Big(\frac{3C_3}{r^3}
 +2(2f_{\xi}+1)\frac{C_6}{r^6}\Big)\delta\theta^2\delta\phi^2. \nonumber
\end{eqnarray}
Similarly, the kinetic term $H_k=-\hbar^2\nabla^2_{\cp r}/m$ gives  
\begin{eqnarray}
	&&H_k^{(0)}(r)= - \frac{\hbar^2}{m r^2} \frac{\partial}{\partial r} \left(r^2 \frac{\partial}{\partial r}\right);  \nonumber\\ 
	&&H_k^{(2)}(r,\delta\theta,\delta\phi)= - \frac{\hbar^2}{m r^2} \left( \frac{\partial^2}{\partial \delta\theta^2} +\frac{\partial^2}{\partial \delta\phi^2} \right); \nonumber\\ 
	&&H_k^{(4)}(r,\delta\theta,\delta\phi)=  - \frac{\hbar^2}{m r^2} \left(-\delta\theta \frac{\partial}{\partial \delta\theta} + \delta\theta^2\frac{\partial^2}{\partial \delta\phi^2} \right).  \nonumber
\end{eqnarray}
Note that we have neglected terms like $\delta\theta^{2n-1} \frac{\partial}{\partial \delta\theta}$ and $\delta\theta^{2n}\frac{\partial^2}{\partial \delta\phi^2} $ with $n\ge2$, which belong to higher-order corrections and contribute very little to the effective potential in small $r$ regime. 

In Eq.~(\ref{HO}), the Hamiltonian $H_k^{(2)}+V^{(2)}$ describes two independent harmonic oscillators in terms of $\delta\theta$ and $\delta\phi$. The zero-point energy is given by $U_{\rm eff}^{(2)}$ in Eq.~(\ref{Ueff2}), and the corresponding wavefunction follows
\begin{eqnarray}
&&\psi_{0}(y;\delta\theta,\delta\phi)
 =\left(\frac{\sqrt{mA}r}{\pi\hbar}\right)^{\frac{1}{2}}
 \left[2f_{\xi}(1+f_{\xi})\right]^{\frac{1}{8}} \nonumber\\
&&\quad\exp\left[-\frac{\sqrt{mA}r}{2\hbar}
 \left(\sqrt{1+f_{\xi}}\delta\theta^2+\sqrt{2f_{\xi}}\delta\phi^2\right)\right] \nonumber
\end{eqnarray}
where  $r=|y|$ and $A= \frac{3C_3}{r^3} + \frac{ 2f_{\xi}C_6 }{r^6}$. 
Based on this, the fourth-order contribution of effective potential in Eq.~(\ref{U4}) can be evaluated as
\begin{eqnarray}
&&U_{\rm eff}^{(4)}(r)=\int d\delta\theta\int d\delta\phi 
 \psi_{0}^*(H_k^{(4)}+V^{(4)})\psi_{0} \nonumber\\
&&\quad=-\frac{\hbar^2}{2mr^2}
 \left(1-\sqrt{\frac{f_{\xi}}{2(1+f_{\xi})}}\right) \nonumber\\
&&\qquad-\frac{\hbar^2}{2mAr^2}
 \left[\frac{3C_3}{r^3}+\frac{(3+13f_{\xi})C_6}{2r^6}\right] \nonumber\\
&&\qquad-\frac{\hbar^2}{2mAr^2}\sqrt{\frac{f_{\xi}}{2(1+f_{\xi})}}
 \left[\frac{3C_3}{r^3}+\frac{2(1+2f_{\xi})C_6}{r^6}\right]. \nonumber
\end{eqnarray}
It shows that  $U_{\rm eff}^{(4)}$ is always negative for all $r$, and in the limit $r\rightarrow 0$ it scales as $-1/r^2$, much smaller than $U_{\rm eff}^{(2)}\sim 1/r^4$ in Eq.~(\ref{Ueff2}). 

\bigskip

{
\subsection*{Heavy-heavy-light system} }

Explicitly, the zeroth- and second-order expansion terms of $H_L({\cp r})$ can be expressed as 
\begin{eqnarray}
	&&H_{L}^{(0)}(r)= - \frac{\hbar^2}{2m r^2} \frac{\partial}{\partial r} \left(r^2 \frac{\partial}{\partial r}\right) \nonumber\\
	&&\quad+ \sum_{\sigma=\pm} \left[
	-\frac{(1+3f_{\xi})C_3}{r_{\sigma}^3} + \frac{ (1-f_{\xi}^2)C_6 }{r_{\sigma}^6}
	\right];  \nonumber\\ 
	&&H_{L}^{(2)}(y,\delta\theta, \delta\phi)=  -\frac{\hbar^2}{2m r^2} \left( \frac{\partial^2}{\partial \delta\theta^2} +\frac{\partial^2}{\partial \delta\phi^2} \right)  \nonumber\\
	&&\quad + \sum_{\sigma=\pm} 
	\frac{ r^2A_{\sigma} }{ r_{\sigma}^2 }
	\left[ (1+f_{\xi}) \delta\theta^2 + 2f_{\xi} \delta\phi^2  \right].
	\label{HL_expansion2}
\end{eqnarray}
In above equations,
$A_{\sigma}= \frac{3C_3}{r_{\sigma}^3} + \frac{ 2f_{\xi}C_6 }{r_{\sigma}^6}$, and we have defined ${\cp r}_{\pm}\equiv {\cp r}\pm {\cp R}/2$ with spherical coordinate $(r_{\pm}, \theta_{\pm}, \phi_{\pm})$, which is associated with angular fluctuations $\{\delta\theta_{\pm}, \delta\phi_{\pm}\}$. Given ${\cp R}$ along $y$, $\{\delta\theta_{\pm}, \delta\phi_{\pm}\}$ can be related to $\{\delta\theta, \delta\phi\}$ via 
\begin{equation}
\delta\theta_{\pm}=r\delta\theta/r_{\pm},\ \ \ \ \delta\phi_{\pm}=y\delta\phi/r_{\pm};
\end{equation}
with $y=r\sin\phi^{(0)}$ the projection of ${\cp r}$ along $y$ direction. 

Note that in obtaining $H_{L}^{(2)}$, we have neglected the angular fluctuations of radial distance $r_{\pm}$ and replaced it with $r_{\pm}=|y\pm R/2|$.
In this way one can recover the correct asymptotic behavior of effective potential when a heavy-light pair come close together ($r_{\pm}\rightarrow 0$). We can see that $H_{L}^{(2)}$ is again composed by two separate harmonic oscillators in terms of   $\delta\theta$ and $\delta\phi$. Its zero-point energy, which gives $U_{\rm L;eff}^{(2)}$, reads 
\begin{equation}
U_{\rm L;eff}^{(2)}(r)= \left( \sqrt{1+f_{\xi}} + \sqrt{2f_{\xi}} \right) 
\sqrt{ \frac{ \hbar^2 }{ 2m } \sum_{\sigma=\pm} 
\frac{ A_{\sigma} }{ r_{\sigma}^2 }}.
\nonumber
\end{equation}  
Besides, the corresponding fourth-order contribution of effective potential can be evaluated as
\begin{eqnarray}
	U_{\rm L;eff}^{(4)}(y)=u_{\rm eff}^{(4)}(r_{+})+ u_{\rm eff}^{(4)}(r_{-})\nonumber
\end{eqnarray} 
with $u_{\rm eff}^{(4)}(r)=U_{\rm eff}^{(4)}(r)/2$.

\bigskip

{
\subsection*{Three identical molecules} }

Specifically, the zeroth- and second-order terms of $H_3$ read 
\begin{eqnarray}
&&H_{3}^{(0)}(r,\rho)
 =-\frac{\hbar^2}{mr^2}\frac{\partial}{\partial r}
 \left(r^2\frac{\partial}{\partial r}\right) \nonumber\\
&&\qquad-\frac{\hbar^2}{m\rho^2}\frac{\partial}{\partial\rho}
 \left(\rho^2\frac{\partial}{\partial\rho}\right) \nonumber\\
&&\qquad-\frac{(1+3f_{\xi})C_3}{r^3}
 +\frac{(1-f_{\xi}^2)C_6}{r^6} \nonumber\\
&&\qquad+\sum_{\sigma=\pm}\left(
 -\frac{(1+3f_{\xi})C_3}{r_{\sigma}^3}
 +\frac{(1-f_{\xi}^2)C_6}{r_{\sigma}^6}\right); \nonumber\\
&&H_{3}^{(2)}(y_r,\delta\theta_r,\delta\phi_r;
 y_{\rho},\delta\theta_{\rho},\delta\phi_{\rho}) \nonumber\\
&&\quad=-\frac{\hbar^2}{mr^2}\left(
 \frac{\partial^2}{\partial\delta\theta_r^2}
 +\frac{\partial^2}{\partial\delta\phi_r^2}\right) \nonumber\\
&&\qquad-\frac{\hbar^2}{m{\rho}^2}\left(
 \frac{\partial^2}{\partial\delta\theta_{\rho}^2}
 +\frac{\partial^2}{\partial\delta\phi_{\rho}^2}\right) \nonumber\\
&&\qquad+A\left[(1+f_{\xi})\delta\theta_r^2
 +2f_{\xi}\delta\phi_r^2\right] \nonumber\\
&&\qquad+\sum_{\sigma=\pm}\frac{A_{\sigma}r^2}{4r_{\sigma}^2}
 \left[(1+f_{\xi})\delta\theta_r^2+2f_{\xi}\delta\phi_r^2\right] \nonumber\\
&&\qquad+\sum_{\sigma=\pm}\frac{3A_{\sigma}{\rho}^2}{4r_{\sigma}^2}
 \left[(1+f_{\xi})\delta\theta_{\rho}^2+2f_{\xi}\delta\phi_{\rho}^2\right] \nonumber\\
&&\qquad+\sum_{\sigma=\pm}\frac{\sqrt{3}A_{\sigma}\sigma}{2r_{\sigma}^2}
 \left[(1+f_{\xi})r\rho\delta\theta_r\delta\theta_{\rho}\right. \nonumber\\
&&\qquad\qquad\left.+2f_{\xi}y_ry_{\rho}\delta\phi\delta\phi_{\rho}\right]. \nonumber
\label{H3_expansion2}
\end{eqnarray}
Here $y_r=r\sin\phi_r^{(0)}$, $y_{\rho}=\rho\sin\phi_{\rho}^{(0)}$ are respectively the projection of ${\cp r}$, $\bm{\rho}$ along $y$ direction; ($r_{\sigma}, \theta_{\sigma}, \phi_{\sigma}$) is the spherical coordinate of ${\cp r}_{\sigma}$ (here $\sigma=\pm$), with angular fluctuations $\{\delta\theta_{\sigma}, \delta\phi_{\sigma}\}$ following 
\begin{eqnarray}
\delta\theta_{\sigma}&=& \frac{r\delta\theta_r+\sigma \sqrt{3}\rho\delta\theta_{\rho}}{2r_{\sigma}}; \nonumber\\
\delta\phi_{\sigma}&=& \frac{y_r\delta\phi_r+\sigma \sqrt{3}y_{\rho}\delta\phi_{\rho}}{2r_{\sigma}}.
\end{eqnarray}
In obtaining above equations, we again neglect the angular fluctuations of radial distance $r_{\pm}$ and directly replace it with $|\frac{y_r\pm \sqrt{3}y_{\rho}}{2}|$. We can see that $H_{3}^{(2)}$ is composed by two independent harmonic oscillators, while within each oscillator the two modes ($\{\delta\theta_r, \delta\theta_{\rho}\}$ or $\{\delta\phi_r, \delta\phi_{\rho}\}$) are entangled with each other. The corresponding zero-point energy $U_{\rm 3;eff}^{(2)}$ can be evaluated following the approach established previously\cite{SWC}.

{\bf Acknowledgements}\\
This work is supported by National Natural Science Foundation of China (12525412, 92476104, 12134015) and Quantum Science and Technology-National Science and Technology Major Project (2024ZD0300600). 

{\bf Author contributions}\\
X.C. conceived the project and supervised the study. H.W and T.S. performed the numerical calculations and analyzed the data. All authors contributed to writing the manuscript.\\

{\bf Competing interests}\\
The authors declare no competing interests.\\

{\it \bf{Data Availability}}\\
The data that support the findings of this article are openly available~\cite{data}.

{\it \bf{Code availability}}\\
The computational codes underlying the figures presented in this study are available from the corresponding author upon request.

\bigskip

{\it \bf{References}}

\onecolumngrid
\clearpage
\setcounter{figure}{0}
\renewcommand{\figurename}{}
\renewcommand{\thefigure}{Supplementary Figure \arabic{figure}}
\renewcommand{\theHfigure}{S\arabic{figure}}
\setcounter{equation}{0}
\renewcommand{\theHequation}{S\arabic{equation}}
\label{supplement-start}
\begin{center}
{\large\bfseries Supplementary Information for\\
Interaction-induced Dimension Reduction for Bound States in Microwave-Shielded\\
Ultracold Molecules\par}
\vspace{0.8em}
Haitian Wang$^{1,2}$, Tingting Shi$^1$, and Xiaoling Cui$^1$\par
\vspace{0.5em}
{\small\itshape
$^1$Beijing National Laboratory for Condensed Matter Physics,\\
Institute of Physics, Chinese Academy of Sciences, Beijing, 100190, China\\
$^2$School of Physical Sciences, University of Chinese Academy of Sciences, Beijing 100049, China\par}
\end{center}

In this supplementary material, we provide more details on the validity of adiabatic approach, the 1D diagrams for various molecular systems and the behavior of bound states near 1D boundary.
We follow the notation and length/energy units defined in the main text.

\bigskip\noindent{\Large\bfseries Supplementary Note 1\quad Validity of adiabatic approach}\par\nobreak\medskip

In the derivation of effective 1D model, we have only kept the ground sate level  ($\nu=0$) and neglected all higher levels ($\nu>0$) of  angular fluctuations. The validity of this treatment can be examined by comparing the binding energy of the system with the characteristic  energy gap for  angular fluctuations. The comparison can be straightforwardly carried out for  tetratomic system. First, we write down the eigen-energy of Eq.~(4) in the main text 
\begin{equation}
E(n_{\theta}, n_{\phi})=(n_{\theta}+1/2)\omega_{\theta}+(n_{\phi}+1/2)\omega_{\phi},
\tag*{Supplementary Equation (1)}
\end{equation}
where $n_{\theta}$ and $n_{\phi}$ are respectively the quantum numbers for angular fluctuations along $z$  and $x$ directions, with eigen-frequencies
\begin{equation}
\omega_{\theta}=2\sqrt{1+f_{\xi}}  
 \sqrt{\frac{\hbar^2}{m} \left( \frac{3C_3}{r^5} + \frac{2 f_{\xi} C_6}{r^8} \right)}, \ \ \ \ \omega_{\phi}= 2\sqrt{2 f_{\xi}} \sqrt{\frac{\hbar^2}{m} \left( \frac{3C_3}{r^5} + \frac{2 f_{\xi} C_6}{r^8} \right)}.
 \tag*{Supplementary Equation (2)}
\end{equation}
The ground state is then $E(0,0)=(\omega_{\theta}+\omega_{\phi})/2$, which gives $U_{\rm eff}^{(2)}$ as shown by Eq.~(5) in the main text. Due to the presence of other terms (such as $H_k^{(n)}$ and $V_k^{(n)}$ expressed in Methods), the ground state can be coupled to higher levels with even numbers of $n_{\theta}$ and $n_{\phi}$. Since $\omega_{\phi}<\omega_{\theta}$, the lowest excitation gap is given by $\Delta E=2\omega_{\phi}$. 

\begin{figure}[h]
\includegraphics[width=8cm]{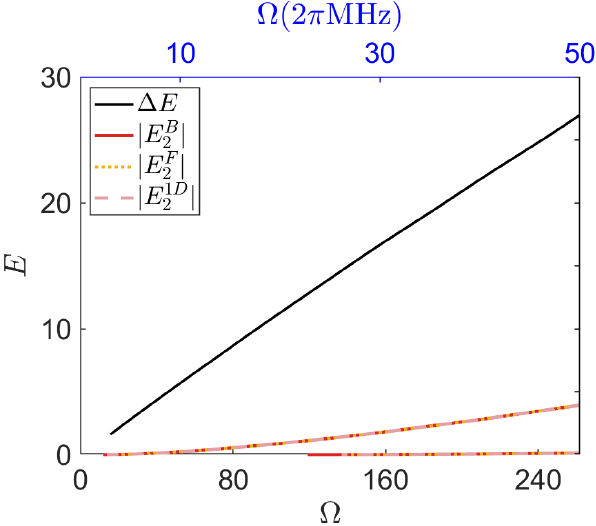}
\caption{Energy comparison between the characteristic transverse gap ($\Delta E=2\omega_{\phi}$) and the binding energy  of tetratomic ground state  ($|E_2^{B/F}|$). Here we take $\xi=\pi/12$. The units of energy and $\Omega$ are $E_u$ and $E_u/\hbar$, respectively.}  \label{fig_adiabatic}
\end{figure}

Given that the gap locally depends on 1D coordinate ($r\equiv|y|$), here we choose a characteristic location $r=r_m$ to evaluate the gap. Specifically, $r_m$ is chosen to be the minimum of the effective 1D potential $V^{(0)}+U_{\rm eff}^{(2)}$,  corresponding to the location for the highest weight of bound state wave-function. In \ref{fig_adiabatic}, we compare  $\Delta E\ (=2\omega_{\phi})$ with the ground state binding energy of tetratomic system at a typical $\xi=\pi/12$ and varying $\Omega$. We can see that for all $\Omega$, $\Delta E$ is substantially larger than the tetratomic binding energies. This validates the ground-mode truncation and the adiabatic approximation employed in the effective 1D description.

In principle, similar comparison can also be carried out for hexatomic system. However, the situation becomes quite complicated due to the presence of more angular fluctuation modes and their mutual entanglement. Here we will not explicitly carry out the comparison. Nevertheless, we can already see that the adiabatic approach works well for heavy-heavy-light system, given the BO potential well predicted by the  effective 1D model as shown in Fig.~5 of the main text. 

\begin{figure}[h]
\includegraphics[width=16cm]{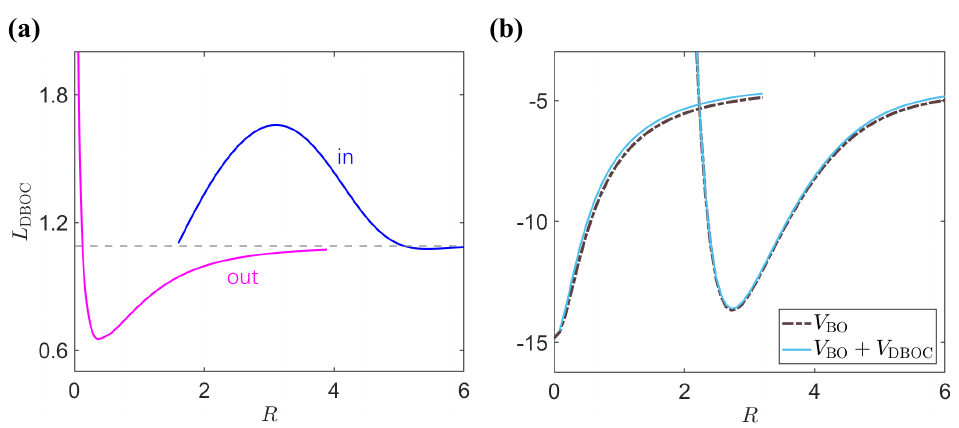}
\caption{(a)Characteristic length ($L_{\mathrm{DBOC}}$) of the diagonal non-adiabatic correction to BO potential. Here we show $L_{\mathrm{DBOC}}$ as a function of $R$ for each BO state as shown in Fig.~5 of the main text. Here 'in' and 'out' respectively denote the light object staying inside and outside of two heavy ones. The horizontal dashed line shows the asymptotic result in $R\rightarrow\infty$ limit, where the light molecule pairs with one heavy molecule to form a tetratomic bound state. (b) Effect of DBOC for a specific mixture of NaLi (light) and NaCs (heavy) molecules, with mass ratio $M/m=156/29$.  The length unit is $l_u$ and the energy unit is $E_u$. }  \label{fig_correction}
\end{figure}

Next we evaluate the correction to  Born-Oppenheimer (BO) potential between two heavy molecules.  
The leading non-adiabatic correction is called the diagonal BO correction (DBOC).
To facilitate its evaluation we utilize the effective 1D description, under which the correction can be written as
\begin{equation}
V_{\mathrm{DBOC}}(R)=-\frac{\hbar^2}{2M}\int d y F^*_0(y,R)\,\frac{\partial^2}{\partial R^2}F_0(y,R).
\tag*{Supplementary Equation (3)}
\end{equation}
Here $F_0(y,R)$ is the 1D wavefunction of light molecule when two heavy molecules are separated (along y direction) by a distance $R$, as solved from the effective 1D Hamiltonian (Eq.~(11) in the main text). Physically, this correction steps from the relative motion of heavy molecules and thus scales as $1/M$, where $M $ is the heavy mass. Then it is convenient to  rewrite $V_{\mathrm{DBOC}}\equiv \hbar^2/(2ML_{\mathrm{DBOC}}^2)$, with characteristic length 
\begin{equation}
L_{\mathrm{DBOC}}=[-\int d y
F_0^*(y,R)\,\frac{\partial^2}{\partial R^2}F_0(y,R)]^{-1/2}.
\tag*{Supplementary Equation (4)}
\end{equation} 
In \ref{fig_correction}(a) we plot $L_{\mathrm{DBOC}}$ as a function of $R$ for two orthogonal BO states considered in the main text.  One can see that $L_{\mathrm{DBOC}}$ is generally comparable to the length unit $l_u$, for the ground state at all $R$ values. Therefore, one expects that  the correction $V_{\mathrm{DBOC}}$ is much smaller than the BO potential evaluated in the main text, given the condition $M\gg m$. We further confirm this in  \ref{fig_correction}(b) by taking a specific molecular mixture with NaLi as light and NaCs as heavy (giving mass ratio  $M/m=156/29$). For this system, one can see that the correction from DBOC is indeed negligibly small across all $R$ regime.  

\clearpage\noindent{\Large\bfseries Supplementary Note 2\quad 1D diagram and bound states near the 1D boundary}\par\nobreak\medskip

\begin{figure}[h]
\includegraphics[width=\textwidth]{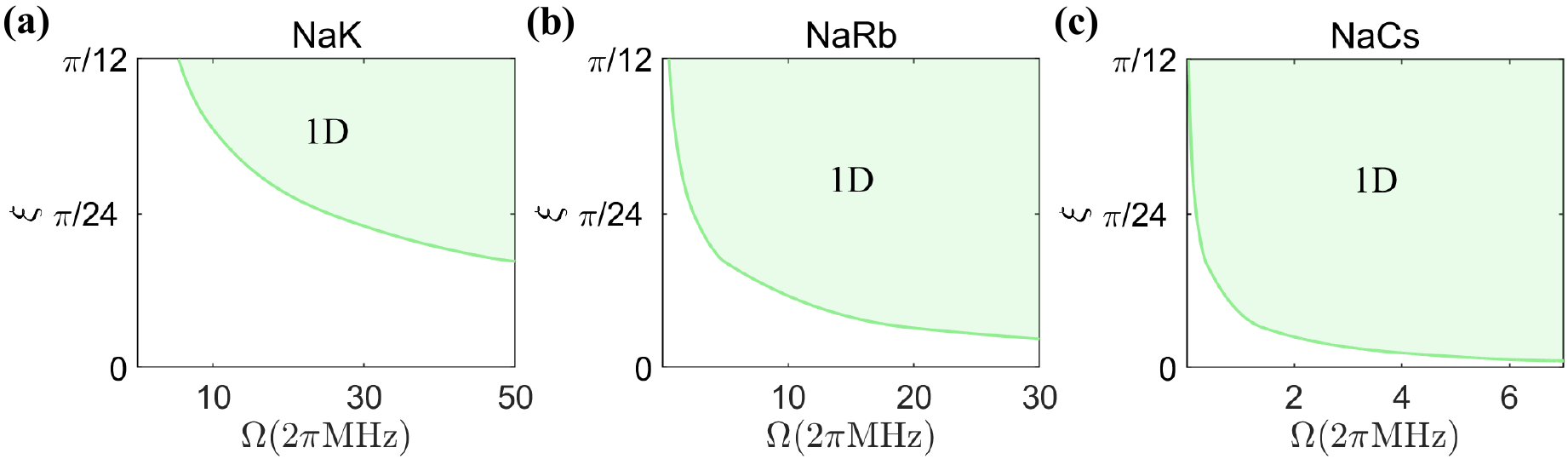}
\caption{ 1D diagrams in ($\Omega, \xi$) plane for NaK (a), NaRb (b) and NaCs (c) systems.  }  \label{fig_1d}
\end{figure}

In Fig.~2(b) of the main text, we have presented a universal 1D diagram in the parameter plane of ($\hbar\Omega/E_u,\ \xi$), with the actual value of $\Omega$ for NaK system shown in the upper axis. Here we will  compare the 1D diagrams in ($\Omega,\ \xi$) plane for different molecular systems including NaK, NaRb and NaCs. Because of larger dipole lengths and heavier masses of NaRb and NaCs, their energy units $E_u$ are much smaller than that of NaK. As a result, at the same $\hbar\Omega/E_u$, NaRb and NaCs  have much smaller $\Omega$ than NaK. It then follows that for the same $\xi$, NaRb and NaCs can enter the effective  1D regime at much smaller $\Omega$ than NaK. 
For instance, at a given $\xi=\pi/24$, NaK, NaRb and NaCs respectively enter the 1D regime at $\Omega=(2\pi)\ 25.32,\ 2.48$ and $0.18 \mathrm{MHz}$, as shown in \ref{fig_1d}. This shows that the 1D description can be more robust for NaRb and NaCs systems with smaller characteristic energy scales. 

\begin{figure}[h]
\includegraphics[width=\textwidth]{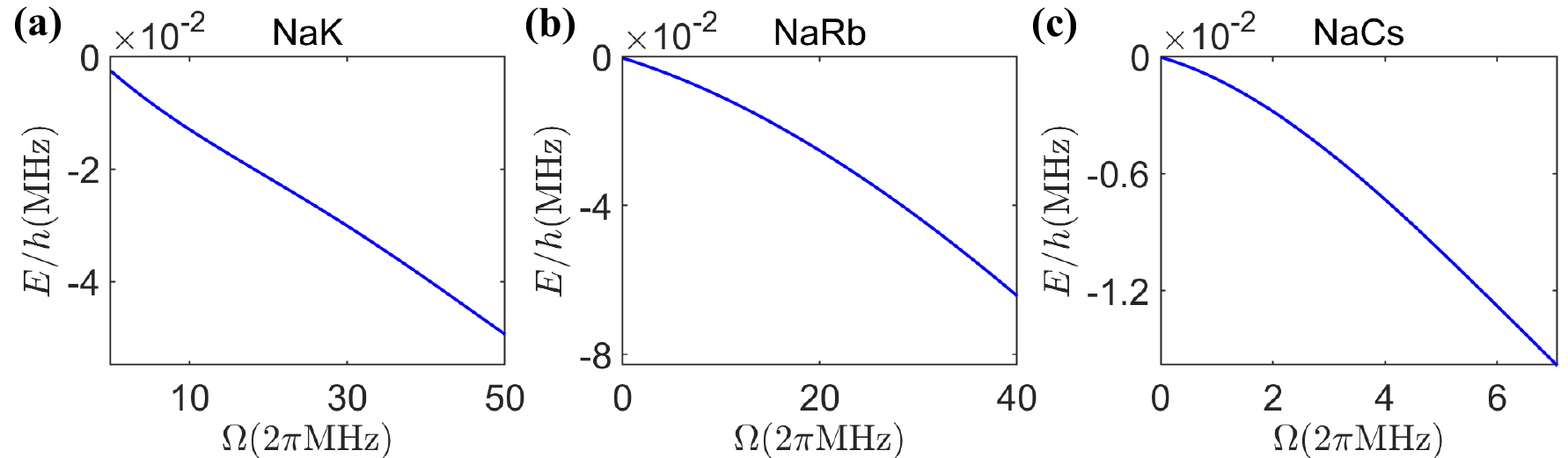}
\caption{Tetratomic binding energies of NaK (a), NaRb (b) and NaCs (c)  at the critical boundary for  1D validity. }  \label{fig_boundary}
\end{figure}

In addition, we would like to emphasize that  the effective 1D model can  apply to quite shallow bound states that are relevant in ultracold gases. To show this, we have computed the binding energy of  tetratomic ground state at the 1D boundary, i.e., identified by the green line in \ref{fig_1d} (or equivalently Fig.~2(b) of the main text). In \ref{fig_boundary}, we show such tetratomic binding energy as a function of $\Omega$ for NaK, NaRb and NaCs systems. One can see that for typical $\Omega$ up to tens of MHz, the binding  energies of all systems are no larger than tens of KHz, well within the reach of current experiment. As shown in a recent NaK experiment [Nature 626, 283 (2024)], the decay rates of  these shallow bound states are sufficiently low to enable their practical detection. Therefore, it is  feasible to explore these shallow bound states described by the effective 1D model, at least near the 1D boundary. Surely for those very deep bound states far from the  boundary, the decay rate could be high enough to make them difficult to explore.

\end{document}